\documentclass[prl,twocolumn,showpacs,superscriptaddress]{revtex4}
\usepackage[xdvi]{graphicx}
\usepackage{dcolumn}
\usepackage{amsmath}

\newcommand{\phic}{\phi_{\rm c}}
\newcommand{\phicc}{\phi_{\rm c}'}
\newcommand{\ep}{\epsilon}
\newcommand{\tauc}{\tau_{\rm c}}
\newcommand{\taucc}{\tau_{\rm c}'}
\newcommand{\epc}{\ep_{\rm c}}
\newcommand{\epcc}{\ep_{\rm c}'}
\newcommand{\bv}[1]{{\boldsymbol #1}}

\begin{document}

\title{History dependence of mechanical properties in granular systems}
\author{Shio Inagaki}
\email{inagaki@chem.scphys.kyoto-u.ac.jp}
\affiliation{Department of Physics, Kyoto University, Kyoto, 606-8502, Japan}
\author{Michio Otuski}
\altaffiliation{Department of Physics and Mathematics, 
Aoyama Gakuin University, Kanagawa, 229-8558, Japan}
\affiliation{Yukawa Institute for Theoretical Physics, Kyoto, 606-8502, Japan}
\author{Shin-ichi Sasa}
\affiliation{Department of Pure and Applied Sciences, University of Tokyo,
Tokyo, 153-8902, Japan}
\date{\today }

\begin{abstract}
We study the history dependence of the mechanical properties of granular 
media by numerical simulations.
We perform a compaction of frictional disk packings in a two-dimensional 
system by controlling the area of the domain with various strain rates. 
We then find the strain rate dependence of 
the critical packing fraction above which the pressure becomes finite.
The observed behavior makes a contrast 
with the well-studied jamming transitions for frictionless disk packings. 
We also observe that the elastic constants of the disk packings 
depend on the strain rate logarithmically. This result provides 
a experimental test for the history dependence of granular
systems.
\end{abstract}

\maketitle


Granular materials have attracted the attentions of scientists 
for a long time \cite{nagel:_granularreview}.
Despite the extensive efforts,
our understanding has not yet been deepened 
enough to construct a fundamental law.
One major difficulty arises from the process dependence of the 
macroscopic behaviors. 
It is rather surprising
to know that the mechanical properties of even static granular 
media cannot be determined from all the information of the constituents.
An example is given by a so-called stress dip, which is a local 
minimum of a vertical pressure distribution just under the apex 
of a sandpile. This stress dip appears by local deposition of grains 
with a point source whereas it does not appear by homogeneous 
deposition with an extended source \cite{vanel99:_memory}. 
Such process dependence is often called as 'memory' which sounds 
mysterious. A sandpile is merely a collection of 
sand grains between which the interaction is rather simple; 
they interact each other only when they are in contact. 
What attracts us is that only granules store information on how 
they are piled up. 
See Refs.~\cite{josserand:_chicagomemory,barrat:_agingmemory,nakahara:_paste}
as other related memory effects.


Granular materials have also been of interest 
as a jamming system \cite{durian:_jamming,dauchot:_jamming}.
As an idealized case of jamming transitions,
the packing problem has been studied intensively with frictionless 
particles 
\cite{ohern:_jamming2,ohern:_jamming1,
MVH:_frictionlessjamming,torquato:_jamming_hardsphere}.
The important result is that 
the pressure of disk packings is not zero above the random close packing
\cite{ohern:_jamming2}.
On the other hand, with frictional particles,
the transition point appears between 
the random loose packing and the random close packing
\cite{MVH:_generaljamming,MVH:_frictionjamming}.
It might be naturally conjectured that the packing 
fraction of the transition  depends on the packing process. 
However, such a dependence has not been well studied.
We aim to get insight into the history dependence of jamming transitions.
Furthermore, we explore a possibility to describe 
the history dependence of mechanical properties of frictional disk packings.


In this Letter, for the purpose of illuminating the nature of the history dependence,
we propose a frictional disk packing in a square box 
without gravity as the simplest system.
We characterize the packing process only by its time scale,
which is precisely given by the inverse of the strain rate 
of the shrinkage in the packing process. By varying this time 
scale in a systematic manner, we attempt to observe the process 
dependence of mechanical properties such as macroscopic elastic constants. 

Here, it should be noted that our numerical experiments are carried out
with identical constituents. In order to examine history dependence 
in an explicit way, it is of the essence to investigate a characteristic 
feature of granular media with persisting with identical constituents.
This makes a contrast  with previous studies investigating 
how macroscopic elastic constants depend on the material properties 
of the particles such as the stiffness, the friction coefficient, 
and the shape  \cite{radjai96:_contactdynamics,
luding01:_2dsimulation, landry02:_3dsimulation}.
The main discovery in the present work is 
an existence of two time scales which are separated over 
two digits. Interestingly, between the two time scales, 
macroscopic elastic properties are  expressed as logarithmic 
functions of the strain rate.

\paragraph{Model:}

We consider $N$  frictional circular disks in a two-dimensional square box. 
We assume that the diameter of  disk $i$, denoted by $d_i$, is either 
$d$ or $d/1.4$ with an equal ratio, and that its mass $m_i$ is given by
$m_i=\pi d_i^2 \rho/4$. The interaction between disks appears 
only when two disks are in contact. Suppose disk $i$ at position
$\bv{r}_{i}$ with velocity $\bv{v}_{i}$ and angular velocity
$\bv{\omega}_{i}$ contacts with disk $j$ at $\bv{r}_{j}$ 
with $\bv{v}_{j}$ and $\bv{\omega}_{j}$. Using the relative 
position $\bv{r}_{ij}=\bv{r}_{i}-\bv{r}_{j}$ and the 
relative velocity $\bv{v}_{ij}=\bv{v}_{i}-\bv{v}_{j}$,
the normal and tangential parts of the relative velocity of contacting points, 
$v^{\rm n}_{ij}$ and  $v^{\rm t}_{ij}$,  are given as 
$v^{\rm n}_{ij}=\bv{v}_{ij}\cdot\bv{n}$ and  $v^{\rm t}_{ij}=
\bv{v}_{ij}\cdot\bv{t}-(d_i\bv{\omega}_{i}+d_j\bv{\omega}_{j})/2$, 
respectively, with the normal unit vector 
$\bv{n}=\bv{r}_{ij}/\left\vert \bv{r}_{ij}\right\vert $ 
and the tangential unit vector $\bv{t}$. Let 
$k_{\rm n}$ and $k_{\rm t}$ be normal and tangential elastic 
constants, $\eta_{\rm n}$ and $\eta_{\rm t}$ be normal and tangential 
viscous coefficients, and $\mu$ be the Coulomb friction coefficient for 
sliding friction \cite{cundall79:_DEM}.
Then, 
the normal force $F_{ij}^{\rm n}$ and the tangential force $F_{ij}^{\rm t}$ 
acting on disk $i$ from disk $j$ are given by
\begin{eqnarray}
F_{ij}^{\rm n} &=& 
-k_{\rm n}(d_{i}/2+d_{j}/2- 
    \left\vert \bv{r}_{ij} \right\vert )
+\eta_{n}v^{\rm n}_{ij},\\
F_{ij}^{\rm t}  & =& 
\min(\left\vert h^{\rm t}_{ij}\right\vert ,\mu\left\vert F_{ij}^{\rm n}
\right\vert ){\rm sign}(h^{\rm t}_{ij})
\end{eqnarray}
with $h^{\rm t}_{ij}=k_{\rm t}u^{\rm t}_{ij}
+\eta_{\rm t}v^{\rm t}_{ij}$. Here, 
$u^{\rm t}_{ij}=\int\nolimits_{t_{0}}^{\rm t}v^{\rm t}_{ij}(t) dt$
is the tangential displacement, where $t_{0}$ is the
time when the disks come into contact. 
The interaction between a disk and the side walls of the box 
is calculated in the same way with the one for the interaction 
between disks.


The parameters in our simulations are converted into dimensionless quantities
so that $d=1$, $k_{\rm n}=1$, and $\rho=1$. Then, the parameter values we 
used are as follows. $d=1$, $\mu=0.4$, $k_{\rm t}=0.1$, and
$\eta_{\rm n}=\eta_{\rm t}=0.35$. Note that the normal 
restitution coefficient is approximately estimated as $0.6$. We restrict 
our investigation to the systems with $N=1000$. 
For a reference to 
laboratory experiments, for example, in the case of disks with maximum 
diameter of $10 {\rm mm}$, 
stiffness of $2.5{\rm MPa}$, thickness of $6 {\rm mm}$, 
and mass density of $1.6 \times {\rm g/cm}^3$, the time unit in our 
simulations corresponds to about $3.1 \times 10^{-4}{\rm s}$.  


The time evolution of the position, velocity and rotation angle of
disk $i$ is described by the equation of motion with using the forces 
described above.  In numerical simulations, we solve a set of equations 
of motion with employing the Adams method with a time step of
$2.2 \times 10^{-2}$. 


\begin{figure}[tbh]
  \begin{tabular}{cc} 
    \begin{minipage}{0.475\columnwidth}
\begin{center}
\rotatebox{-90}{\includegraphics[width=3.8cm]{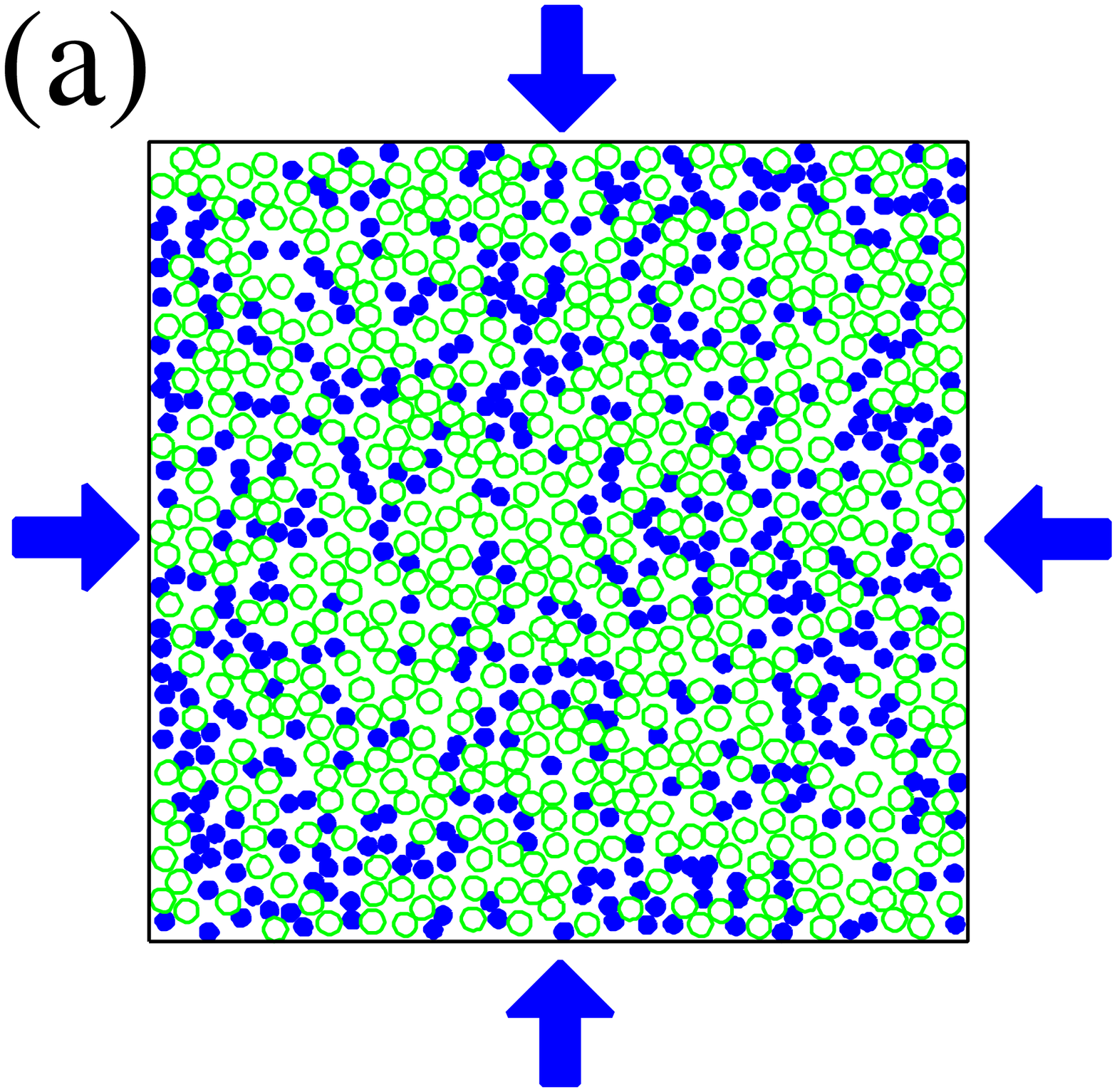}}
\end{center}
    \end{minipage} &
    \begin{minipage}{0.475\columnwidth}
\begin{center}
\includegraphics[width=3.8cm]{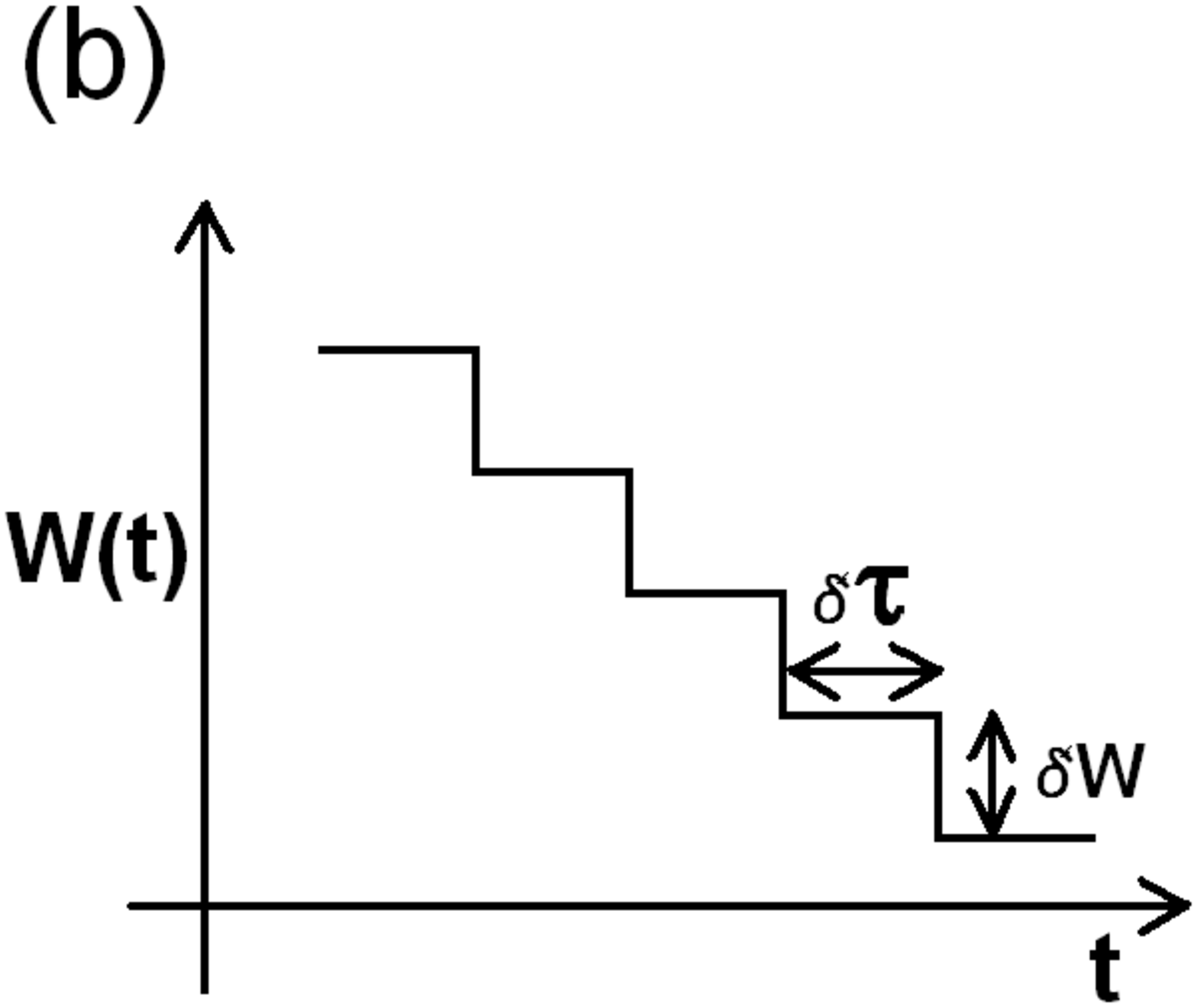}
\end{center}
    \end{minipage}
  \end{tabular}
\caption{Illustration of the packing process. (a) Initial configuration: 
Disks are placed in a box. Open (green online) and filled (blue online)
circles represent the particles 
with  diameters of $1.0$ and $1.0/1.4$, respectively. (b) Schematic 
diagram of the motion of walls: Width of a box, $W(t)$, is reduced 
as a function of time. The box is shrunken isotropically with 
various strain rates.
}%
\label{fig:packingprocess}%
\end{figure}

%
%

As an initial state at $t=0$, disks are scattered in a  square box 
with no contact. The width of the box $W(t=0)$
is determined so that the initial packing fraction is to be $0.5$.
(See Fig.~\ref{fig:packingprocess} (a).)
We then carry out 
the following compaction procedure with moving all sides of the 
box simultaneously toward the center of the box. Let $\delta W$ be a 
small value. The width of the box is reduced with $\delta W$ 
at intervals of time $\delta \tau$. The time dependence of
the width  of the box $W$ is illustrated in Fig.~\ref{fig:packingprocess} (b). 
At the moment the width of the box is reduced with the strain, 
centers of disks are shifted with 
an affine transformation with an equal amount of the strain of the box.

%
%

We investigate cases with several values of $\delta \tau$ with 
$\delta W=5.0 \times 10^{-2}$ fixed. We confirmed that our claim
in this Letter is 
independent of the value of $\delta W$ if it is sufficiently small. 
We stop moving the walls at certain time $t=\tau$ and we wait 
until the total kinetic energy of disks becomes smaller than 
$10^{-14}$ so that we can assume it to be in a static state. 
This compaction process is characterized by the strain rate 
measured from the initial state:
\begin{equation}
\dot{\epsilon}
=\frac{W(t=0)-W(t=\tau)}{W(t=0)\tau}.
\end{equation}
Note that since the first contact occurs immediately after 
we start a compaction,
we assume the initial configuration as a reference state to 
determine the strain rate.
Below we will regard  the strain rate $\dot{\epsilon}$ as the 
control parameter of the problem we consider.   

\paragraph{Evidence of the process dependence:}

We first study the process dependence of the final pressure
with the packing fraction fixed. In this study,
we stop the shrinkage of the walls when the packing fraction 
reaches  a prescribed value. 

\begin{figure}[tbh]
\begin{center}
\includegraphics[width=6cm]{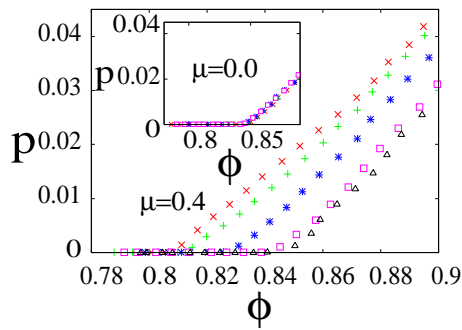}
\end{center}
\caption{Pressure $p$ versus  packing fraction $\phi$
for the system with  $\mu=0.4$ and various values of
strain rate $\dot \ep$, $1.42\times 10^{-3}$ (Cross symbol),
$3.56\times10^{-4}$ (Plus symbol), 
$1.78\times10^{-4}$ (Asterisk symbol)), and
$4.45\times10^{-5}$ (Square symbol)).
For comparison, the result in the case 
$\mu=0.0$ and the strain rate of $4.45\times 10^{-5}$ 
is superposed as the triangle symbols. 
Inset: Results in the case $\mu=0.0$, with various strain rates.
The symbols coincide with those in the main frame, but all 
the symbols are on one curve. The average is taken over 20 samples.
}%
\label{fig:phi_fri0p4}%
\end{figure}

For comparison, we consider the case $\mu=0.0$. 
As shown in the inset of Fig.~\ref{fig:phi_fri0p4}, the results 
with four different strain 
rates yield one pressure curve as a function of packing fraction. 
Here, the packing fraction of the system is measured
in the domain away from the boundary at the distance of $2d$
in order to remove boundary effects.
We observe that the pressure starts increasing at a critical packing 
fraction $\phic\simeq 0.84$, which corresponds to
the so-called jamming transition point \cite{torquato:_jamming_hardsphere}.
It is a remarkable property that the critical packing fraction
for frictionless particles is determined uniquely 
irrespective of packing processes. 

\begin{figure}[tbh]
\begin{center}
\includegraphics[width=6cm]{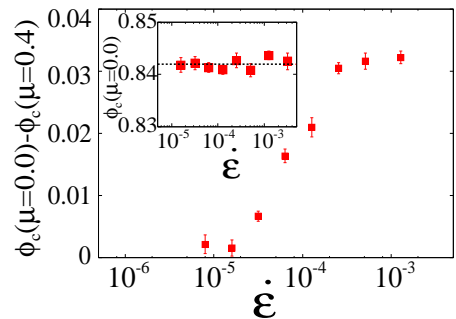}
\end{center}
\caption{
Deviation of the critical packing fraction $\phic(\mu=0.4)$ from
$\phic(\mu=0.0)$ as a function of the strain rate of the shrinkage,
where $\phic(\mu=0.0)=0.842$ is assumed. Inset: Strain rate independence
of critical packing fraction $\phic$ with $\mu=0.0$. 
}%
\label{fig:phip_mu}%
\end{figure}

In contrast, under the existence of the tangential friction
with the friction coefficient of $\mu=0.4$, the pressure curve
strongly depends on the packing process as shown in Fig.~\ref{fig:phi_fri0p4}. 
In particular, as we pack slower, the pressure curve approaches 
the one in the case $\mu=0.0$ (displayed by the triangle symbols). 
In order to extract a quantitative relation of the process dependence, 
we measured  the critical packing fraction  $\phic$ as a 
function of $\dot{\epsilon}$. As shown in Fig.~\ref{fig:phip_mu},
our result suggests 
\begin{equation}
\phic(\mu=0.4, \dot \ep)-\phic(\mu=0) 
\simeq   -\alpha_\phi \log \frac{\dot \ep}{\dot \epc}
\label{numerics}
\end{equation}
in the regime 
$\dot \epc \le  \dot \ep \le \dot \epcc$ 
with $\dot \epc \ll \dot \epcc$,
where $\dot \epc $ and $ \dot \epcc$ correspond
to the inverse of two cross-over time scales
$\tauc$ and $\taucc$, respectively.

We shall present a phenomenological argument to understand Eq.
(\ref{numerics}). Let us fix $p$ of the final state. When the 
shrinkage is faster than the change of contact network and the force 
balance is realized in this network, the packing fraction in the 
final state does not seriously depend on the strain rate.
This provides the first cross-over time scale $\taucc$. 
When the shrinkage is slower than this time scale, the disks have enough time 
to search for a more dense network. 
The searching time $\tau$ for a configuration
with $\phi$ might be proportional to the number of force balance 
configurations $\Sigma$ for a given $\phi$.  Here, 
due to  the combinatorial complexity of force balance states,
$\Sigma \simeq \exp(aN)$ in the large $N$ limit with a constant 
$a$ that characterizes the number of local configurations satisfying 
force balance conditions. Since it is reasonable to assume that
$a$ can be expanded in $\phi$ near 
the packing fraction $\phicc$ at which the 
shrinkage rate dependence starts increasing,  
$\Sigma$ depends 
on $\phi$ in an exponential manner. The consideration leads us to 
\begin{equation}
\tau \simeq \taucc \exp( A (\phi-\phicc)),
\end{equation}
where  $A$ is a constant. 
This tendency ceases when the packing fraction $\phic$ is close 
to $\phic(\mu=0)$, which provides the time scale $\tauc$ as
$\tauc =\taucc \exp( A (\phic(\mu=0)-\phicc))$
in the case $p \simeq 0$.
Since the searching time $\tau$  is related to $1/\dot \ep$ linearly,
we obtain Eq. (\ref{numerics}). 


\paragraph{Experimental method:}

In many laboratory experiments, the packing fraction is not easily 
measurable. We thus propose a method by which the process dependence 
can be detected without observing the packing fraction. The key idea 
here is to focus our attention on mechanical properties of the packing. 
Concretely, we fix the pressure $p$ in the packing. That is, we stop 
the compaction when the final pressure becomes an prescribed value. 
In the present study, we consider the case $p=0.025$. 
As is seen from Fig.~\ref{fig:phi_fri0p4}, 
the packing fraction might depend on the strain rate of the shrinkage, 
and therefore it is naturally expected that Young's  modulus $E$ 
and Poisson's ration $\nu$ of the packings also depend on 
the strain rate $\dot \ep$.

These elastic constants of disk packings can be measured
by performing a bi-axial compression test \cite{inagaki:_enudef}
with the final states of the packing process. 
We measure increments of stress, $\delta\sigma_{ij}$, of 
the system when the strain, $\delta\epsilon_{yy}=0.01$ is given 
while $\delta\epsilon_{x}=0.0$. Then, assuming the linear elastic 
theory, we estimate the values of $E$ and $\nu$. Note that we  
carefully performed the compression test so that the quasi-static 
condition is satisfied during the compression.

\begin{figure}[tbh]
\begin{center}
\rotatebox{-90}{\includegraphics[height=6cm]{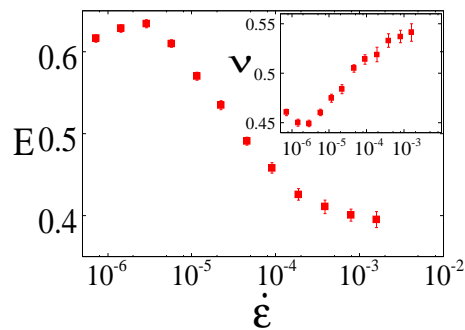}}
\end{center}
\caption{Young's modulus $E$ and Poisson's ratio $\nu$ (inset)
as  functions of the strain rate  $\dot{\epsilon}$.
Young's modulus $E=1$ corresponds to the dimensionless 
normal stiffness of the constituent itself. 
$40$ samples are taken for each $\tau$.
}%
\label{fig:e_tp_log}%
\end{figure}

The results are summarized in Fig.~\ref{fig:e_tp_log}.
We find that the largest Young's modulus is about $1.54$ times larger 
than the smallest one. This fact clearly 
indicates  that the elastic properties depend on the packing 
process despite  the material properties of constituents are 
identical. Furthermore, 
the shapes of curves in Fig.~\ref{fig:e_tp_log} suggests
\begin{align}
E  &  = E_0 - \alpha_E \log(\dot \epsilon),\label{eq:E_log}  \\
\nu &  =\nu_0 + \alpha_{\nu} \log(\dot \epsilon) \label{eq:nu_log},
\end{align}
in the regime 
$\dot \epc \le  \dot \ep \le \dot \epcc$.

Here we wish to remark on a possibility of realizing laboratory 
experiments with our idea. For example, for the experimental system 
we consider in the paragraph {\it Model}, the regime $10^{-6} \le 
\dot{\epsilon} \le 10^{-4}$ can be realized by reducing the size of 
a box from $46 {\rm cm} $ to $38 {\rm cm}$ with  compaction time 
$\tau^{\rm exp}$ satisfying  $0.3 ({\rm s}) \le \tau^{\rm exp} \le 30 
({\rm s})$. This operation is accessible in laboratory  experiments. 
Our simple setting will provide us a great advantage 
to investigate history dependence of granular media.
We expect that the logarithmic dependence of $E$ and $\nu$ on $\dot \ep$
will be examined by laboratory experiments.

\paragraph{Conclusion and discussion:}

To conclude,  we have found the history dependence of the jamming transition
points. For this phenomenon, the tangential friction is found to be 
essential. Furthermore, we have present a clear evidence for the 
packing process dependence of the elastic properties of disk packings 
with identical constituents.

\begin{figure}[ptbh]
\begin{center}
\rotatebox{-90}{\includegraphics[height=5cm]{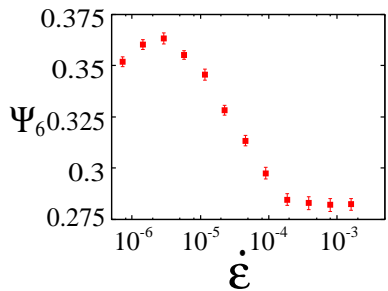}}
\end{center}
\caption{Bond-orientational order parameter $\psi_{6}$ as a function of
the strain rate  $\dot{\epsilon}$.}%
\label{fig:psi6}%
\end{figure}

The most important question left to be answered is to uncover the
universal nature of the logarithmic dependence of $E$ and $\nu$ on 
$\dot \ep$. As  discussed above, the simple argument suggests 
the logarithmic dependence of the packing fraction on $\dot \epsilon$
under the condition that the pressure $p$ in the final state is fixed.
Then, one naively conjectures that the state with  larger packing
fraction with $p$ fixed corresponds to a state closer to crystal in 
random packings. 
In order to confirm this conjecture, 
we measure the extent of crystallization of the packings with
the six-fold order parameter of interbond angle 
defined by
\begin{equation}
\psi_{6}=\left\vert \frac{1}{N_{\rm bond}}%
\sum_{i} \sum_{\langle jk \rangle } \exp(6i\theta^{i}_{jk})
\right\vert,
\label{eq:psi6}
\end{equation}
where $\langle  j k \rangle$ represents a pair of disks 
which are in contact with disk $i$ next to each other,
$\theta^{i}_{jk}$ is the interbond angle between bond $ij$ and $ik$,
and $N_{\rm bond}$ is the total number of such interbond angles.
Here, bond $ij$ means the line segment connecting the centers 
of disk $i$ and disk $j$. Note that 
$\sum_{\langle jk \rangle } \theta^{i}_{jk}=2 \pi$ for each $i$.
As shown in Fig.~\ref{fig:psi6}, the extent of crystallization 
decreases rapidly between the two cross-over time scales.
Therefore, the understanding of the tendency to crystallization 
in slower operations might be a heart of future problems.
We will study it from several viewpoints.

The authors thank  Hisao Hayakawa and Hiroki Ohta 
for fruitful discussions.  S. I. acknowledges 
Ken-ichi Yoshikawa for his helpful comments and warm encouragement.
This work was supported by grants Nos. 
18GS0421 (S. I.), 19840027 (S. I.), and 19540394 (S. S.)
from the Ministry of Education, Science, Sports and 
Culture of Japan.  M. O. thanks 
the Yukawa Foundation for the financial support.


\end{document}